\documentclass[apj,twocolumn]{openjournal}
\usepackage{amsmath}
\usepackage[dvipsnames]{xcolor} 
\usepackage[breaklinks,colorlinks,citecolor=blue,urlcolor=blue,linkcolor=blue]{hyperref}
\usepackage{orcidlink}
\usepackage{graphicx}
\usepackage{amssymb}
\usepackage[dvipsnames]{xcolor}
\usepackage{subfigure}

\definecolor{tofixred}{RGB}{184,13,84}

\definecolor{editpurple}{RGB}{134,65,255}

\begin{document}

\title{The Kinematic Properties of T\.ZO Candidate HV 11417 with Gaia DR3}
\shorttitle{HV 11417 Gaia DR3}

\author{Anna J. G. O'Grady\,\orcidlink{0000-0002-7296-6547}$^{1}$}

\affiliation{$^1$McWilliams Center for Cosmology and Astrophysics, Department of Physics, Carnegie Mellon University, Pittsburgh, PA 15213, USA}

\email{Corresponding author: aogrady@andrew.cmu.edu}

\begin{abstract}
   HV 11417 is a candidate Thorne-\.Zytkow Object, a red supergiant with a neutron star core, located within the Small Magellanic Cloud (SMC). Previous studies have questioned, using \textit{Gaia} DR2 data, whether HV 11417 was truly located at the distance of the SMC or was instead a foreground star. However, the proper motion measurement uncertainties for HV 11417 in DR2 were high. In this work, we use \textit{Gaia} DR3 data to show that HV 11417 is very likely to be a true member of the SMC. We further analyze the kinematics of HV 11417 relative to its local environment, and compare it to populations of massive and evolved stars in the SMC. We find HV 11417 has a local transverse velocity of $52\pm15$ km/s, and thus qualifies as a runaway star (v$_\mathrm{loc}\geq$ 30 km/s). This runaway classification does not conclusively prove its nature as a T\.ZO, particularly given results from recent T\.ZO models, but does indicate that HV 11417 experienced a kinematic disruption in its evolution. 
\end{abstract}

\maketitle
\section{Introduction}\label{sec:intro}

Thorne-\.Zytkow Objects (T\.ZOs), first theorized by \citet{Thorne.K.1975.TZOprime,Thorne.K.1977.TZOstructure}, are a theoretical and intriguing avenue of binary evolution, wherein a neutron star (NS) is engulfed by the envelope of a companion red supergiant (RSG) and merges with its core, forming a single object. Significant work has gone into formalizing the physics of their formation \citep{Taam.R.1978.TZOXRBperiods,Ray.A.1987.TZODynam,Podsiadlowski.P.1995.TZOEvolution,Fryer.C.1996.NSRapidInfallBH,Everson.R.2024.TZOFormation,HutchinsonSmith.T.2024.TZOFormation,Nathaniel.K.2025.TZOPopSynth,Williams.L.2025.TZOMSForm}, evolution \citep{Cannon.R.1992.TZOStrucEvo,Cannon.R.1993.TZOStructure,Biehle.G.1994.TZOObservational,Podsiadlowski.P.1995.TZOEvolution,Farmer.R.2023.TZOModernModels}, final fates \citep{Podsiadlowski.P.1995.TZOEvolution,Liu.W.2015.LongPXraySourceTZO,Moriya.T.2018.TZOExplosions,Moryia.T.2021.TZOExplosions}, and observational properties -- see \citet{OGrady.A.2025.TZOReview} for a review.

To date, the best studied T\.ZO candidate is HV 2112 \citep{Levesque.E.2014.HV2112disc}, a cool, luminous star in the Small Magellanic Cloud (SMC), though its true nature has been debated \citep{Tout.C.2014.HV2112SAGB,Maccarone.T.2016.TZOHighMotion,Beasor.E.2018.HV2112AGB,McMillan.P.2018.HV2112Gaia,O'Grady.A.2020.superAGBidentification,OGrady.A.2023.sAGBTZOAnalysis,Farmer.R.2023.TZOModernModels}. A second T\.ZO candidate, HV 11417, also located in the SMC, was identified as a candidate by \citet{Beasor.E.2018.HV2112AGB}. The candidacy of this star has also been debated, both from the perspective of new T\.ZO models \citep{Farmer.R.2023.TZOModernModels}, but also because its membership within the SMC was brought into question by \citet{O'Grady.A.2020.superAGBidentification}. Using \textit{Gaia} DR2 \citep{Gaia.Collab.2018.GaiaDR2} kinematics, \citet{O'Grady.A.2020.superAGBidentification} found that the proper motion of HV 11417 was marginally inconsistent with the proper motions of SMC stars, suggesting it was instead a foreground star and thus far too dim to be a T\.ZO. However, measurement uncertainties within \textit{Gaia} DR2, especially for the $\mu_\alpha$ of HV 11417 ($-0.307\pm0.359$), were very large.

The 3rd Data Release of \textit{Gaia} \citep{Gaia.2023.DR3} has significantly reduced proper motion measurement uncertainties, allowing for a re-analysis of the proposed location and overall kinematic properties of HV 11417. A kinematic analysis may help constrain the true identity of HV 11417, since T\.ZO formation necessitates a supernova explosion, which may impart a kick to the system. 

In this paper, we use \textit{Gaia} DR3 to analyze the kinematic properties of HV 11417 in the context of a possible T\.ZO identity and compared to massive stars in the SMC in general. In \S\ref{sec:location} we determine whether or not HV 11417 is located within the SMC. Then, in \S\ref{sec:local_env} we compare the motion of HV 11417 to other stars in its local environment, and to other populations of massive and evolved stars in the SMC. Finally, we discuss our results in \S\ref{sec:discussion} and conclude in \S\ref{sec:conclu}.

\section{Confirming SMC Membership}\label{sec:location}

We adopt a distance of 62.44 kpc \citep{Graczyk.D.2020.SMCDist} and mean proper motions of $\mu_\alpha = 0.797$, $\mu_\delta = -1.22$ \citep{Gaia.2021.EDR3Clouds} for the SMC. To determine if the proper motions of HV 11417 from \textit{Gaia} DR3 are consistent with membership in the SMC, we apply the method previously used by \citet{O'Grady.A.2020.superAGBidentification,OGrady.A.2023.sAGBTZOAnalysis,OGrady.A.2024.YSGBinI}, modeled after a procedure described by \cite{Gaia2018}, for removing likely foreground sources. A given star is considered a likely foreground source if i) its parallax over parallax error $\pi/\sigma_\pi > 4$, and/or ii) its proper motions fall outside the 99.5 percentile range of the proper motion phase space of a covariance matrix describing the proper motion distribution of $\sim$ 1,000,000 highly probable SMC members (described in detail in \S2.5 of \citealt{O'Grady.A.2020.superAGBidentification}). Specifically, a source is a likely foreground star if, 

\begin{equation}\label{eq:chi2}
    \chi^{2}_\mathrm{SMC} = ({\mu}-\overrightarrow{\mu}_{\textrm{med}} )^{T}\mathrm{C}_{*}^{-1}({\mu}-\overrightarrow{\mu}_{\textrm{med}}) > 10.6,
\end{equation}

\noindent where $\overrightarrow{\mu}$ is the proper motion of the star, and $C_{*}$ and $\overrightarrow{\mu}_{\textrm{med}}$ are the covariance matrix (including measurement uncertainties) and median proper motions, respectively, of the highly probable SMC stars.

The proper motion values of HV 11417, shown in Table \ref{tab:kinematics}, are consistent with membership in the SMC ($\chi^{2}_{\mathrm{SMC}}$ = 0.57 $\ll$ 10.6), and its parallax over parallax error is -2.3, well below 4. For reference, the \textit{Gaia} DR2 proper motion values for HV 11417 were $\mu_\alpha = -0.307\pm0.359$ and $\mu_\delta = -1.769\pm0.21$. The extreme $\mu_\alpha$ value led to the previous uncertainty in the true location of HV 11417.

\begin{table*}
    \centering
    \begin{tabular}{c||c|c|c|c|c|c|c|c|c|c}
        \hline
        Star & RA & Dec & $\mu_\alpha$ & $\mu_\delta$ & $\chi^{2}_{\mathrm{SMC}}$\footnote{See Equation \ref{eq:chi2}} & $\epsilon$ (Sig)\footnote{Astrometric Excess Noise (Significance)} & RUWE & CDF\footnote{Percentage denotes how much of the cumulative density function of local stars is interior to the star} & Speed\footnote{Relative to the mean speed of local stars} & $\sigma_\mathrm{Local}$\footnote{Standard deviation of the proper motion of local stars} \\
         & degree & degree & (mas yr$^{-1}$) & (mas yr$^{-1}$) & & & & & km/s & km/s \\ \hline \hline
         HV 11417 & 15.20070 & $-$72.85056 & 0.59$\pm$0.05 & $-$1.28$\pm$0.05 & 0.57 & 0.44 (34) & 1.01 & 53.5\% & 52$\pm$15 & 38 \\
         HV 2112 & 17.51607 & $-$72.61461 & 1.04$\pm$0.03 & $-$1.27$\pm$0.03 & 2.36 & 0.24 (44) & 1.14 & 56.6\% & 54$\pm$9 & 30  \\
         M2002 1631 & 10.55134 & $-$73.38681 & 0.51$\pm$0.03 & $-$1.16$\pm$0.03 & 1.32 & 0.09 (2) & 0.98 & 33.6\% & 33$\pm$16 & 39  \\
         M2002 15271 & 12.50778 & $-$72.19072 & 0.45$\pm$0.03 & $-$1.26$\pm$0.03 & 1.78 & 0.22 (30) & 0.97 & 75.0\% & 67$\pm$14 & 35  \\
         M2002 30492 & 13.63938 & $-$73.6837 & 0.48$\pm$0.03 & $-$1.29$\pm$0.02 & 1.82 & 0.08 (2) & 0.95 & 61.6\% & 60$\pm$13 & 40  \\
         M2002 37502 & 14.14036 & $-$72.21111 & 0.31$\pm$0.03 & $-$1.33$\pm$0.03 & 5.19 & 0.24 (20) & 1.58 & 95.0\% & 125$\pm$14 & 43 
         
    \end{tabular}
    \caption{The \textit{Gaia} DR3 and derived kinematic properties of HV 11417 and RIOTS4 comparison stars. All values are taken or calculated from \textit{Gaia} DR3 data, other than the speeds of the RIOTS4 comparison stars, taken from Table 1 of \citet{Phillips.G.2024.RIOTS4.4}}
    \label{tab:kinematics}
\end{table*}

\section{Kinematic Properties Relative to Local Environment}\label{sec:local_env}

\subsection{Local proper motion \& transverse velocity of HV 11417}

Here we compare the kinematics of HV 11417 relative to other stars in its local environment. We again follow the method of \citet{OGrady.A.2023.sAGBTZOAnalysis} to analyze the kinematics of HV 11417 compared to its local environment. We query Gaia DR3 for all sources within 5$^\prime$ of HV 11417, which corresponds to a radius of $\sim 90$ pc at the distance of the SMC. Foreground stars are filtered out using the same method as \S\ref{sec:location}, and we additionally add a magnitude restriction of $G \leq 18$ mag, following both \citet{OGrady.A.2023.sAGBTZOAnalysis} and \citet{Phillips.G.2024.RIOTS4.4}. 

\begin{figure*}
    \centering
    \includegraphics[width=\textwidth]{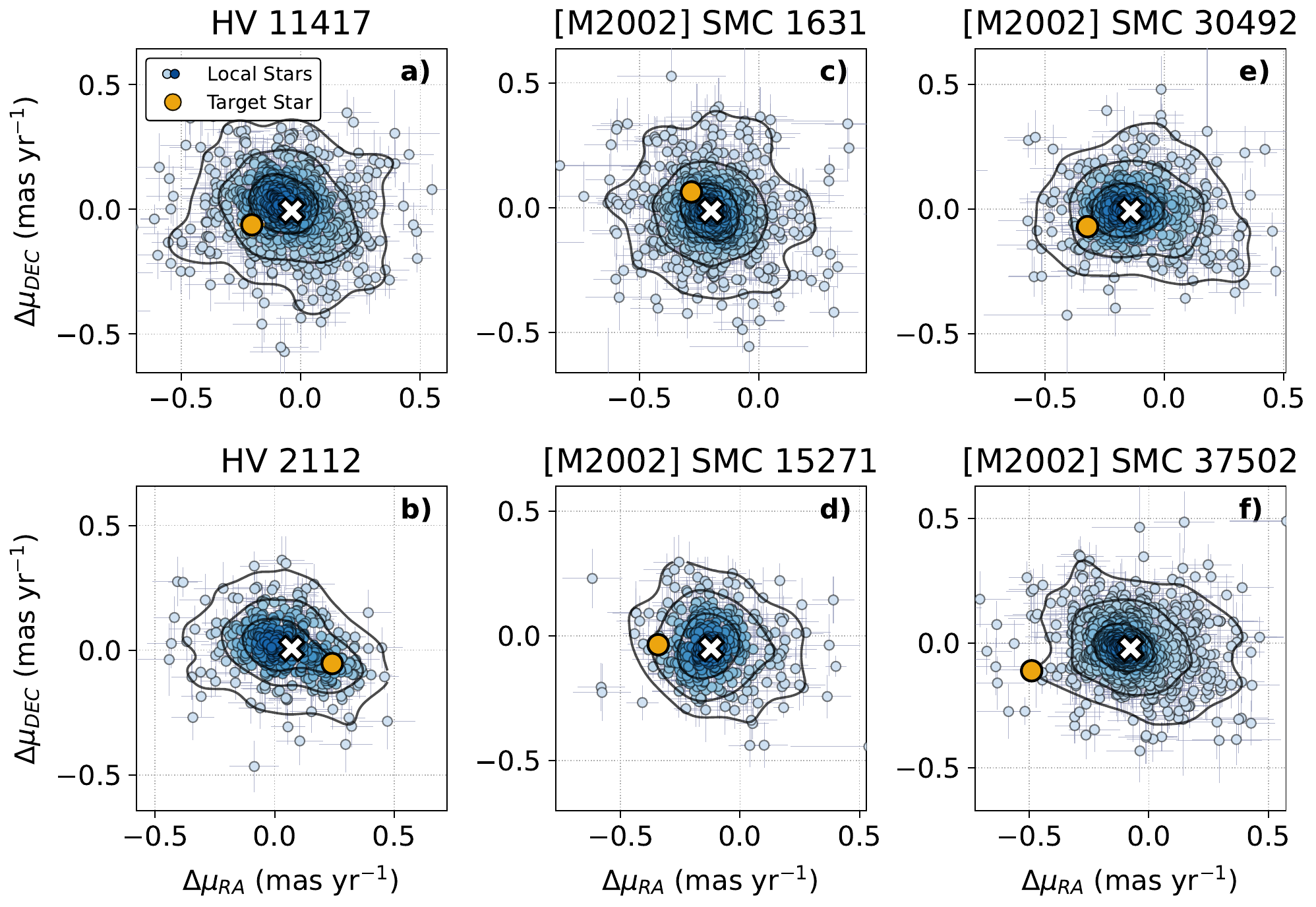}
    \caption{The two-dimensional proper motions of the local environments of HV 11417 (panel a) and comparison stars (panels b-f). The star of interest is a large gold circle, and all stars within 5$^\prime$ that are likely SMC sources are small blue circles, where darker blue indicates a higher density. A white cross indicates the mean proper motion for each group. Contours indicate the 1-, 2-, and 3-$\sigma$ densities of the local stars. The mean proper motion of the SMC has been subtracted, and the star of interest is not included in the local mean or density contour determination.}
    \label{fig:twod_kin}
\end{figure*}

In Figure \ref{fig:twod_kin}, the proper motions of HV 11417 (gold circle) and the surrounding stars (blue circles) are plotted in the top left panel (a). The white cross indicates the mean proper motion of the local stars, weighted by measurement uncertainties and \emph{excluding} HV 11417, so that if HV 11417's motion is peculiar relative to its local environment it will not bias the mean. Black lines indicate the 1-, 2-, and 3-$\sigma$ contours of the local density distribution. HV 11417 falls just outside the 1-$\sigma$ contour.

\begin{figure}
    \centering
    \includegraphics[width=\columnwidth]{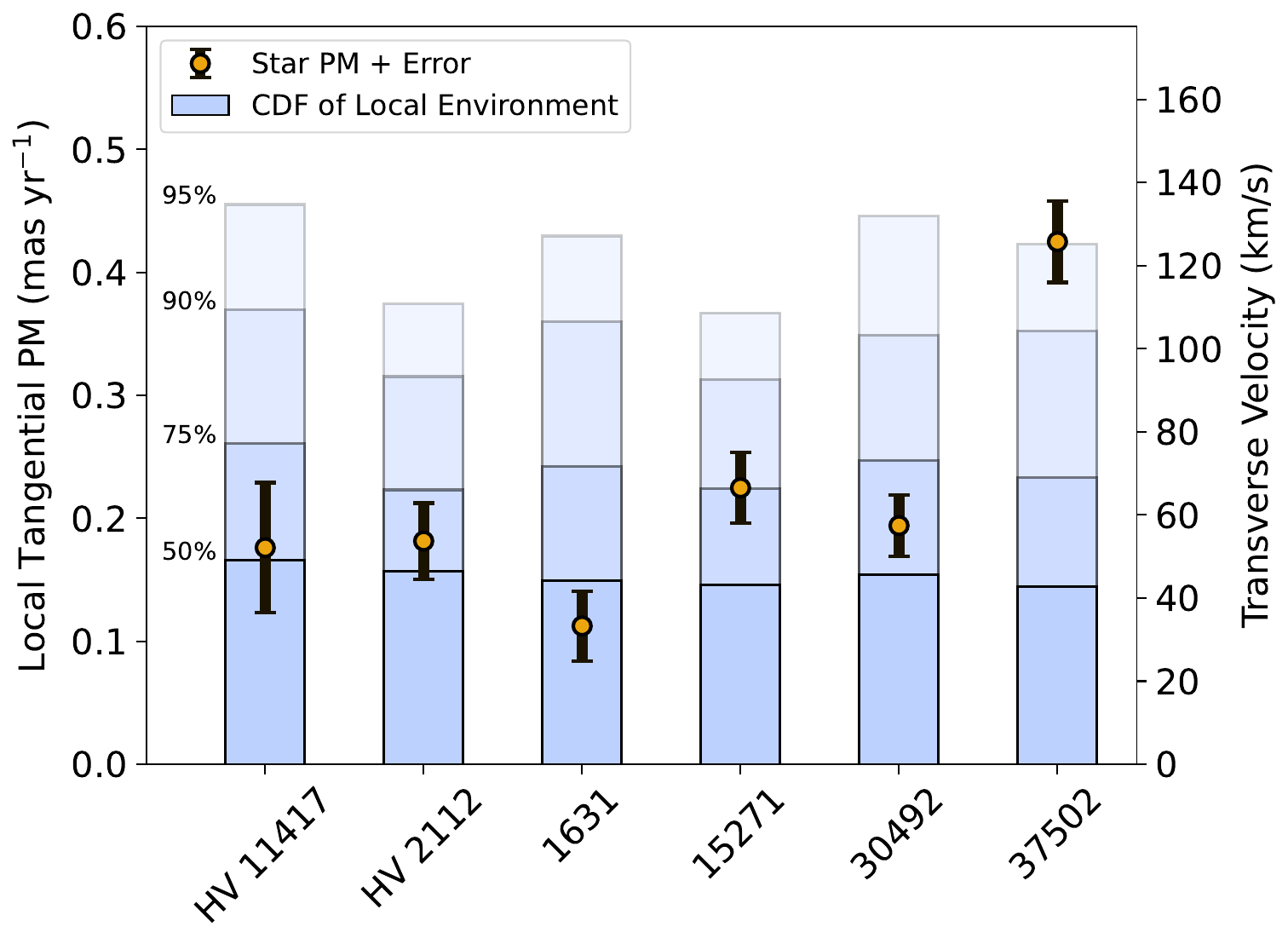}
    \caption{The tangential proper motion (left axis) and transverse velocity at the distance of the SMC (right axis) of HV 11417 and comparison stars, shown as gold circles with black error bars. The segmented blue shaded bars indicate the 50\%, 75\%, 90\%, and 95\% percentiles of the tangential proper motion cumulative density function of all likely SMC stars within 5$^\prime$ of each star of interest. The mean proper motion of the SMC \emph{and} the weighted mean of each locale has been subtracted, allowing the proper motion of each star of interest to be compared to the average dispersion of their local environment.
    }
    \label{fig:oned_kin}
\end{figure}

We further investigate the total tangential proper motion and transverse velocity of HV 11417 relative to its local environment in Figure \ref{fig:oned_kin}, in the leftmost bar. HV 11417 has a transverse velocity, relative to nearby stars, of 52$\pm$15 km/s. This is faster than $\sim$ 54\% of the surrounding stars, and is consistent with being a runaway star (under the criterion of v$_\mathrm{loc}\geq$ 30 km/s; \citealt{Blaauw.A.1961.RunawayStars,Lamb.J.2016.RIOTS4.1}).

\subsection{Comparison to other stars}

We now compare the kinematic properties of HV 11417 to the T\.ZO candidate HV 2112, and to populations of massive and evolved stars in the SMC. 

\medskip
\noindent \textit{HV 2112:} The best studied T\.ZO candidate, the kinematic properties of HV 2112 were previously analyzed by \citet{OGrady.A.2023.sAGBTZOAnalysis}. We show HV 2112 in panel (b) of Figure \ref{fig:twod_kin} and as the second bar in Figure \ref{fig:oned_kin}. HV 2112 displays very similar properties to HV 11417 -- both are located near the 1-$\sigma$ boundary of the 2D proper motion distribution of their local environments, both have transverse velocities of $\sim$ 50 km/s, and for both that velocity is faster than $\sim 50\%$ of nearby stars. 

\medskip 
\noindent \textit{Runaway OB stars:} To identify appropriate comparison stars we use The Runaways and Isolated O-Type Star Spectroscopic Survey of the SMC (RIOTS4; \citealt{Lamb.J.2016.RIOTS4.1,Oey.S.2018.RIOTS4.2,Dorigo.J.2020.RIOTS4.3,Phillips.G.2024.RIOTS4.4}), which found that $\sim 65\%$ of field OB stars in the SMC are runaways (with velocities $\geq$ 30 km/s) and that the dynamical ejection scenario (DES, i.e. from the star's birth cluster) dominates over the binary supernova scenario (BSS, induced by the natal kick of the explosion). We chose 4 stars from Table 1 of \citet{Phillips.G.2024.RIOTS4.4} to plot as comparative examples -- M2002 1631, 15271, 30492, and 37502 (see Table \ref{tab:kinematics}). These stars were selected according to the following criteria: i) We consider only stars within the `main body' of the SMC, not within the SMC Wing, as our covariance matrix for removing likely foreground sources is based on likely SMC stars, and the Wing has substantially different motions \citep{Oey.S.2018.RIOTS4.2}; ii) We consider only stars identified as emission line and/or high mass X-ray binaries in RIOTS4, as these classes are most likely to have experienced a BSS kick rather than a DES ejection \citep{Dallas.M.2022.OBePostSN,Phillips.G.2024.RIOTS4.4}; and iii) We chose stars at a range of transverse velocities corresponding to the broad range of expected T\.ZO velocities, plus one very fast star ($\sim 125$ km/s, M2002 37502) that is likely the result of a combined DES+BSS ejection \citep{Phillips.G.2024.RIOTS4.4} as an extreme example. These stars are plotted in panels (c)-(f) in Figure \ref{fig:twod_kin} and the rest of the bars in Figure \ref{fig:oned_kin}. We can see that HV 11417, as well as HV 2112, indeed meet the criteria to be runaway stars, though the uncertainties for HV 11417 are large. 

When comparing to the RIOTS4 sample of likely BSS kicked stars, 43\% have v$_\mathrm{loc}\geq$ 30 km/s, and 15\% have v$_\mathrm{loc}\geq$ 50 km/s. When compared against the complete RIOTS4 sample (of all field OB/OBe stars), 21\% have v$_\mathrm{loc}\geq$ 30 km/s and 7\% have v$_\mathrm{loc}\geq$ 50 km/s.

\medskip
\noindent \textit{Runaway AGBs/RSGs:} We compare as well to the bulk kinematic properties of RSGs, a progenitor population for T\.ZOs, and asymptotic giant branch (AGB) stars, the most likely stellar classification for HV 11417 if it is not a T\.ZO \citep{OGrady.A.2023.sAGBTZOAnalysis}. For RSGs we use SMC sources (removing SMC Wing stars) from the catalog of \citet{Ren.Y.2021.RSGsLocalGroup}. For the AGB sample, we select oxygen-rich AGB stars (O-AGB), as HV 11417 does not demonstrate the high mass loss rate associated with carbon-rich AGB stars\footnote{From the relations of \citet{Matsuura.M.2013.GlobalGasDustSMC} and 2MASS, Spitzer, and WISE photometry, the mass loss rate of HV 11417 is $\sim 1\times10^{-6}$M$_\odot$yr$^-1$.}, in the SMC from the Spitzer-SAGE catalog \citep{Boyer.M.2011.SAGE.MC.Photom}. For each RSG and AGB star, we remove likely foreground stars and perform the same kinematic analysis as above. We find that 33/55\% of RSGs/O-AGB stars have local transverse velocities $\geq$ 30 km/s, 12/23\% of RSGs/O-AGB stars have local transverse velocities $\geq$ 50 km/s, and that 13/26\% of RSGs/O-AGB stars have local transverse velocities faster than 50\% of other local stars.

\section{Discussion}\label{sec:discussion}

\subsection{Kinematic properties of HV 11417}

The improved measurement uncertainties of \textit{Gaia} DR3 enable a confident placement of HV 11417 at the distance of the SMC. Its transverse velocity, relative to the mean velocities of nearby stars, is $\sim 52\pm15$ km/s, consistent with the criterion for being a runaway star ($\geq$ 30 km/s). The measurement uncertainty for HV 11417 is still high (0.05 mas yr$^-1$ in $\mu_\alpha$ and $\mu_\delta$ compared to 0.03 mas yr$^-1$ for HV 2112 and other comparison stars, Table \ref{tab:kinematics}). HV 11417 has a higher transverse velocity than $\sim 54\%$ of its local environment, though we also note the relatively high velocity dispersion (38 km/s) of those nearby stars.

We consider this runaway classification of HV 11417 in the context of massive and evolved star populations in the SMC. The RIOTS4 survey \citep{Lamb.J.2016.RIOTS4.1} finds that 65\% of field OB stars are runaways, and \citet{Phillips.G.2024.RIOTS4.4} find with \textit{Gaia} DR3 specifically that 46\% of field B stars have v$_\mathrm{loc}\geq$ 30 km/s (and 20\% show v$_\mathrm{loc}\geq$ 50 km/s). Further, in this work we find that 55\% of O-rich AGB stars and 33\% of RSGs in the SMC have v$_\mathrm{loc}\geq$ 30 km/s (and 23\% and 12\% have v$_\mathrm{loc}\geq$ 50 km/s).

We note that HV 11417 has a high astrometric excess noise parameter ($\epsilon$) of 0.44 with a high $\epsilon$ significance of 34, indicating excess scatter in \textit{Gaia's} position measurements, but a Renormalized Unit Weight Error (RUWE) of 1.01, which indicates that the single-star astrometric model fits well. While photometric variability can cause elevated $\epsilon$ values \citep{Chiavassa.A.2022.GaiaAEN,Gandhi.P.2022.AEN}, the distance of the SMC is too large for the apparent photocenter of even a highly variable star to wobble enough to produce 0.44 mas of excess noise. Binarity can cause high $\epsilon$ values, but the excellent single-star astrometric fit (RUWE=1.01) indicates that HV 11417 is likely a single star, though the presence of a companion at a very large separation or in a face-on configuration would not be captured by the RUWE parameter \citep{Belokurov.V.2020.GaiaUnresComp}. Most likely, the high and significant value of $\epsilon$ stems from the crowded nature of SMC fields \citep{Vasiliev.E.2018.GaiaDR2LMC,Fabricius.C.2021.EDR3Val}.

\subsection{Implications for classification}

If HV 11417 is a T\.ZO, its kinematic properties may reflect its formation history, as described in \S2.1 of \citet{OGrady.A.2023.sAGBTZOAnalysis}. \citet{Leonard.P.1994.TZOKick} predict the median velocity for T\.ZOs formed by the natal kick launching the NS into the secondary to be $\sim 75$ km/s. Assuming generally isotropic velocities, this corresponds to a local transverse velocity of  v$_{\mathrm{loc}}\approx 61$ km/s. Alternatively, T\.ZOs formed from the common envelope spiral-in after a HMXB phase (\citealt{Taam.R.1978.TZOXRBperiods}, but see also \citealt{Papish.O.2015.TZOEjectsEnvelope,HutchinsonSmith.T.2024.TZOFormation,Everson.R.2024.TZOFormation}) could have velocities in the range of $\sim$ 10-80 km/s \citep{vandenHeuvel2000}, corresponding to v$_{\mathrm{loc}}\approx$ 8-65 km/s. 

The local transverse velocity of HV 11417 of $\sim 52\pm15$ km/s is thus consistent with either of these formation scenarios, though this alone is not enough to distinguish it from evolved stars unbound through non-T\.ZO scenarios (see Figure 5 of \citealt{Nathaniel.K.2025.TZOPopSynth}). Combined with the luminosity (log(L/L$_\odot)\sim$ 4.92) and elemental abundance anomalies uncovered by \citet{Beasor.E.2018.HV2112AGB} (particularly the enhanced Rb; although \citet{Beasor.E.2018.HV2112AGB} did not detect enhanced Li, see \S5.7.3 of \citet{OGrady.A.2023.sAGBTZOAnalysis} for discussion) the properties of HV 11417 indeed align with predictions from earlier T\.ZO models \citep{Thorne.K.1975.TZOprime,Thorne.K.1977.TZOstructure,Cannon.R.1992.TZOStrucEvo,Cannon.R.1993.TZOStructure}. 

HV 11417 and the T\.ZO candidate HV 2112 share many kinematic similarities. They both have local transverse velocities greater than 50 km/s, both have velocities faster than 50\% of local stars, and both lie near the 1-$\sigma$ density contour in the 2D proper motion distribution. HV 2112 also has similar values of $\epsilon$ (0.24), $\epsilon$ significance (44), and RUWE (1.14) as HV 11417. Both are also variable stars \citep{Soszynski.I.2011.OGLELPVsinSMC}, though HV 2112 demonstrates a much higher optical amplitude ($\Delta$M$_\mathrm{V}\sim$ 4.0) than HV 11417 ($\Delta$M$_\mathrm{V}\sim$ 2.4).  

However, \citet{Farmer.R.2023.TZOModernModels} recently created new models of T\.ZOs with \texttt{MESA}, finding several differences compared to the older models of \citet{Cannon.R.1992.TZOStrucEvo,Cannon.R.1993.TZOStructure}. The luminosity, metallicity, and variability period ($\sim$ 1090 day fundamental mode) of HV 11417 cannot be consistently matched by any of the \citet{Farmer.R.2023.TZOModernModels} models, thus \citet{Farmer.R.2023.TZOModernModels} rule HV 11417 out as a T\.ZO candidate. We do note that the structure of the \citet{Farmer.R.2023.TZOModernModels} models is only computed down to the point where the envelope becomes fully convective (colloquially known as the `knee'), and instead inject energy at the base of the convective envelope in lieu of modeling the radiative zone and neutron star atmosphere. This may have consequences for the predicted nucleosynthetic yields of \citet{Farmer.R.2023.TZOModernModels} if material from below the `knee' can be mixed into the convective envelope. As well, invoking a low NS mass of $\lesssim 1.2$M$_{\odot}$ allows the models to reach the luminosity of HV 11417 (see Appendix 3 and Figure A1c of \citealt{Farmer.R.2023.TZOModernModels}).

If HV 11417 is not a T\.ZO, its luminosity, pulsation properties, and element abundance anomalies are best explained by a massive or super-AGB star identity \citep{Doherty.C.2017.SAGBStarsECSNE}, similar to that of HV 2112 and other similar stars explored in \citet{O'Grady.A.2020.superAGBidentification,OGrady.A.2023.sAGBTZOAnalysis}. The runaway velocity of HV 11417 is not at odds with such a classification, as even a $\sim$ 5M$_\odot$ AGB star would have a B-type main sequence progenitor, and as discussed above, nearly half of B-type main sequence stars in the SMC are runaways \citep{Phillips.G.2024.RIOTS4.4}.

\section{Conclusion}\label{sec:conclu}


In this paper we re-analyzed the kinematic properties of the T\.ZO candidate HV 11417 using \textit{Gaia} DR3 proper motions. We confirmed that HV 11417 is highly likely to be a member of the SMC, and found that its transverse velocity, relative to its local environment, meets the criterion of a runaway star, with v$_\mathrm{loc}\sim 52\pm15$ km/s. This does not conclusively distinguish between possible T\.ZO formation channels, nor whether HV 11417 is a T\.ZO or some other class of cool evolved star. Given the high fraction of OB main sequence runaways (65\%), and even the high fraction of runaway luminous, cool, evolved stars (33/55\% for RSGs/AGBs), HV 11417 is not kinematically unusual. Its runaway velocity does indicate that it underwent some kinematically disruptive event; whether this was a dynamical ejection, an unbinding supernova kick, or a T\.ZO formation specifically remains uncertain. This work, along with many others, demonstrates the exceptional wealth of information within the \textit{Gaia} data. We encourage further work on the kinematics of massive, evolved stars within the Magellanic Clouds, especially at the population level, enabled by \textit{Gaia} data.

\section*{Acknowledgments}

The author thanks Maria Drout, Katie Breivik, and Adiv Paradise for useful conversations. The author thanks the anonymous referee for helpful suggestions that have improved this manuscript. The author gratefully acknowledges support from the McWilliams Postdoctoral Fellowship in the McWilliams Center for Cosmology and Astrophysics at Carnegie Mellon University. 

Assistance in formatting the `Local Stars' element in the legend of Figure \ref{fig:twod_kin} and proofreading the manuscript for spelling and grammar errors was performed with Claude, Sonnet/Opus 4.5 Model, Anthropic. 

\textit{Software:} This work made use of the following software packages: \texttt{astropy} \citep{astropy:2013,astropy:2018,astropy:2022}, \texttt{Jupyter} \citep{2007CSE.....9c..21P,kluyver2016jupyter}, \texttt{matplotlib} \citep{Hunter:2007}, \texttt{numpy} \citep{numpy}, \texttt{pandas} \citep{mckinney-proc-scipy-2010,pandas_15597513}, \texttt{pyia} \citep{pyia}, \texttt{python} \citep{python}, and \texttt{scipy} \citep{2020SciPy-NMeth,scipy_14880408}.

This research has made use of the Astrophysics Data System, funded by NASA under Cooperative Agreement 80NSSC21M00561; the SIMBAD database and the cross-match service operated at CDS, Strasbourg, France \citep{SIMBAD.Wenger.M.2000}; and data from the European Space Agency (ESA) mission
{\it Gaia} (\url{https://www.cosmos.esa.int/gaia}), processed by the {\it Gaia}
Data Processing and Analysis Consortium (DPAC,
\url{https://www.cosmos.esa.int/web/gaia/dpac/consortium}).

Software citation information aggregated using \texttt{\href{https://www.tomwagg.com/software-citation-station/}{The Software Citation Station}} \citep{software-citation-station-paper,software-citation-station-zenodo}.

\clearpage

\bibliographystyle{aasjournal}
\bibliography{thesis_bib,software}

\end{document}